\documentclass[aps,preprint,nofootinbib]{revtex4}

\usepackage[utf8]{inputenc}
\usepackage{color}
\usepackage{amsmath,amssymb}
\usepackage{amsfonts}
\usepackage{verbatim}
\usepackage{makeidx}
\usepackage{hyperref}
\usepackage{slashed}
\usepackage[vcentermath]{youngtab}
\usepackage{float}
\usepackage{booktabs}
\usepackage{graphicx}
\usepackage{mathrsfs}
\usepackage{bm}
\usepackage{braket}
\usepackage{url}
\usepackage{etoolbox}
\apptocmd{\sloppy}{\hbadness 10000\relax}{}{}
\allowdisplaybreaks[3]
\def\arraystretch{1.0}

 \newcommand{\be}{\begin{equation}}
   \newcommand{\ee}{\end{equation}}
     \newcommand{\bea}{\begin{eqnarray}}
   \newcommand{\eea}{\end{eqnarray}}

\usepackage[normalem]{ulem}  
\usepackage{color}
\renewcommand\sout{\bgroup \color{red} \ULdepth=-.5ex \ULset}
\begin{document}

\title{Search for   $bb\bar{b}\bar{b}$ tetraquark decays in 4 muons, 
 $B^{+} B^{-}$, $B^0 \bar{B}^0$ and  $B_s^0 \bar{B}_s^0$ channels at  LHC}

\author{C. Becchi$^1$}
%\address{INFN, Sezione di Genova, via Dodecaneso 33, 16146 Genova, Italy}
\author{A. Giachino$^1$}
%\address{INFN, Sezione di Genova, via Dodecaneso 33, 16146 Genova, Italy}
\author{L. Maiani$^2$}
%\address{T. D. Lee Institute, Shanghai Jiao Tong   University, Shanghai, 200240, China}
%\address{Dipartimento di Fisica and INFN,  Sapienza  Universit\`a di Roma, Piazzale Aldo Moro 2, I-00185 Roma, Italy}
\author{E. Santopinto$^1$}
%\address{INFN, Sezione di Genova, via Dodecaneso 33, 16146 Genova, Italy}

\affiliation{$^1$Istituto Nazionale di Fisica Nucleare (INFN), Sezione di
Genova, via Dodecaneso 33, 16146 Genova, Italy
}

\affiliation{$^2$T. D. Lee Institute, Shanghai Jiao Tong   University, Shanghai, 200240, China}
\affiliation{$^2$Dipartimento di Fisica and INFN,  Sapienza  Universit\`a di Roma, Piazzale Aldo Moro 2, I-00185 Roma, Italy}

\begin{abstract}
We perform a quantitative analysis of the $bb\bar{b}\bar{b}$ tetraquark decays into hidden- and open-bottom mesons and 
calculate, for the first time,  the  $bb\bar{b}\bar{b}$ tetraquark total decay width.
On the basis  of our results, we propose the $bb\bar{b}\bar{b} \to B^{+} B^{-} (B^0 \bar{B}^0)  (B_s^0 \bar{B}_s^0) \to l^{+} l^{-}+\text{X}$ decays  as the most  suitable  channels to observe the  $bb\bar{b}\bar{b}$ tetraquark states, since the  calculated two-lepton cross section upper limit, $\simeq  39 $ fb, is so large as to be potentially detectable with the 2018 LHCb sensitivity,  
 paving the way to the observation of  the  $bb\bar{b}\bar{b}$ tetraquark in the future LHCb upgrade. The $4\mu$ signal for the ground state, $J^{PC}=0^{++}$, is likely to be too small even for the upgraded LHCb, but it may not be hopeless for the $J^{PC}=2^{++}$ fully-bottom state.
\end{abstract}
%\pacs{12.39.Hg,12.39.Fe,14.20.-c,21.30.Fe}
  \maketitle

\section{\bf{Introduction}}
\label{sec1}
The hypothetical existence of hadronic states with more than the minimal quark content ($q\bar{q}$ or $qqq$) was proposed by Gell-Mann in 1964  \cite{GellMann:1964nj} and Zweig \cite{Zweig:1964}, followed by the construction
of a quantitative model for two quarks plus two antiquarks by Jaffe \cite{Jaffe:1976ig}, to describe the lightest scalar mesons.
Recent years have seen  considerable growth in the observations of  four valence quark  states that cannot be included in the well-known systematics of mesons made up  of quark-antiquarks, $Z(4248)$, $Z(4430)$, etc.. Similar particles have also been found in the bottom sector, $Z_b(10610)$ and $Z_b(10650)$, observed by the Belle collaboration \cite{Bondar}  (see  \cite{Ali:2019roi} or \cite{Esposito:2016noz} for recent reviews).

The first  predictions of a fully-bottom, $b b \bar{b} \bar{b} $ tetraquark below the $2{\Upsilon}$ threshold were made in Refs. \cite{Heller:1985cb,Berezhnoy:2011xn}, 
and were supported by more recent contributions~\cite{Wu:2016vtq,Chen:2016jxd,Bai:2016int,Wang:2017jtz,Richard:2017vry,Anwar:2017toa}.
Theoretically, $J^{PC}=0^{++}$ is expected for the $b b \bar{b} \bar{b}$ ground-state. 
%For completeness, we shall consider also the case of $J^{PC}=2$ tetraquark.

In 2018, LHCb performed  a search for $bb\bar{b}\bar{b}$  decaying into four-muons in the mass range 17.5-20 GeV, but no 
significant excess was found, leading to the 95\% CL upper limit~\cite{LHCb-PAPER-2018-027} 
\be
\sigma(pp\to {\cal T})\times B({\cal T}\to \Upsilon(1S) \mu^{+} \mu^{-} ) \times B(\Upsilon(1S)\to \mu^{+} \mu^{-} )< 20~{\rm fb}\label{uplhcb}.
\ee
% While  tetraquark  mass spectroscopy has been addressed in many works 
%\cite{Heller:1985cb,Berezhnoy:2011xn,Wu:2016vtq,Chen:2016jxd,Bai:2016int,Wang:2017jtz,Richard:2017vry,Anwar:2017toa},   there are no quantitative predictions for the $bb\bar{b}\bar{b}$ 
%Preliminary estimates of the fully-bottom tetraquark  decay widths into leptons has been given in Refs. \cite{Esposito:2018cwh} and \cite{Karliner:2016zzc}. 

Ref. \cite{Esposito:2018cwh} estimates the $J^{PC}=0^{++}$, fully-bottom tetraquark decay width into  $\Upsilon \mu^{+} \mu^{-} $ to be in the range: $10^{-3}-10$ MeV  \cite{Esposito:2018cwh}. Ref. \cite{Karliner:2016zzc} gives a total decay width of 1.2 MeV, a partial decay width into four-leptons in the range: $2.4 \cdot10^{-3}-2.4  \cdot 10^{-7}$ MeV with a branching ratio in the range:  $2 \cdot10^{-3}-2  \cdot 10^{-7}$.
%{\color{red}

In this letter, assuming the mass of the $J^{PC}=0^{++}$ fully-bottom  tetraquark to lie below the $2\eta_b$ threshold, we present for the first time a calculation of decay widths and branching ratios of the main, hidden- and open-bottom channels. Our results are as follows.
%\begin{itemize}

The total width is expressed as: $\Gamma({\cal T}(J=0^{++})=7.7~{\rm MeV}\cdot\xi$, where $ \xi $  is the ratio of the overlap probabilities of the annihilating $b \bar b$ pairs in $\cal T$ and $\Upsilon$, respectively. Following Ref.~\cite{Anwar:2017toa}.we estimate:
%{\color{red} {\bf CHECK CHECK} references}
\be
\xi_{\rm th}= \frac{|\Psi_{\cal T}(0)|^2}{ |\Psi_{\Upsilon}(0)|^2} \sim 1.1^{+0.9}_{-0.5}  \label{bestratio} \; {\bf \to}\;
\Gamma({\cal T})=8.5~{\rm MeV} ~{\rm (best ~guess)} %\label{bestguess1}
\ee
 Decay rates are all proportional to $\xi$ so that the branching fractions are uniquely determined; these are  reported in Table~\ref{uno}.  
 In particular, we find 
\bea
&& B({\cal T}\to 4\mu)=7.2\cdot 10^{-7}. \label{B4mufin} 
\eea

The result \eqref{B4mufin}  is not far from the lower limit of the range in~\cite{Karliner:2016zzc}, the reason being that the total tetraquark width is in the order of the $\eta_b$ rather than of the $\Upsilon(1S)$ width, which gives a ${\cal O}(10^{-3})$ suppression.

 With \eqref{B4mufin}, we obtain a realistic estimate of 
%The $bb\bar{b}\bar{b}$ branching fraction into four muons allows, for the first time, to estimate 
the  cross section for $p+p \to{\cal T}\to\Upsilon+ \mu^+\mu^-\to 4 \mu$. Our result  is about 350  times lower than the 95\% CL upper limit quoted in \eqref{uplhcb}.
On the other hand, the  calculated  cross section for the tetraquark strong decays  into two $B_q \bar B_q$ mesons, ($q=u,d,s,c$) is large enough, see Tab.~\ref{tab:width},  to be potentially detectable in the future LHCb upgrade~\cite{Bediaga:2018lhg}. 
%\end{itemize}

We repeated the calculation for the $J^{PC}=2^{++}$, fully-bottom tetraquark, assuming it lay below the $2\eta_b$ threshold. The $J=2$ tetraquark is produced in $p+p$ collisions with a statistical factor of $5$ with respect to the spin $0$ state;  the decay  ${\cal T} \to \eta_b+{\rm light~ hadrons}$ is suppressed. However,
annihilation into two vector mesons $M^*_q \bar M^*_q$ takes place  at a greater rate than for $0^{++}$. Branching fractions of $J^{PC}=2^{++}$ are listed in Tab.~\ref{uno} and, with \eqref{bestratio}, 
\be
 \Gamma({\cal T}(J=2^{++})= 12~{\rm MeV} ~{\rm (best ~guess)} \label{bestguess2}
\ee
.
 %@@@@@@@@@@@@@@@@@@@@@@
  \begin{table}[htb!]
\centering
    \begin{tabular}{|c|c|c|c|c|c|c|}
     \hline
{\footnotesize $[bb][\bar b \bar b]$}&{\footnotesize $\eta_b$+ any} & {\footnotesize $B_q \bar B_q$ ($q=u, d, s, c$)} & {\footnotesize $ B^*_q \bar B^*_q$}& {\footnotesize $\Upsilon_b$+ any} & {\footnotesize $\Upsilon_b+\mu^+\mu^-$} &  {\footnotesize $4\mu$} \\ \hline 
{\footnotesize $J^{PC}=0^{++}$}&{\footnotesize $0.65$}&{\footnotesize $0.022$}  &{\footnotesize $0.066$}&{\footnotesize $1.2\cdot 10^{-3}$} & {\footnotesize $2.9\cdot 10^{-5}$} & {\footnotesize $7.2\cdot 10^{-7}$}   \\ \hline
{\footnotesize $J^{PC}=2^{++}$}&{\footnotesize $0$}&{\footnotesize $0$}  &{\footnotesize $0.25$}&{\footnotesize $3.4\cdot 10^{-3}$} & {\footnotesize $8.3\cdot 10^{-5}$} & {\footnotesize $20\cdot 10^{-7}$}\\ \hline 
\end{tabular}
 \caption{\footnotesize {Branching ratios of $J^{PC}=0^{++}$ and $2^{++}$ fully-bottom tetraquarks, masses below $2\eta_b$ threshold, assuming $S$-wave decay.}}
\label{uno}
\end{table}  
  %@@@@@@@@@@@@@@@@@@@@

\section{\bf{Details of the calculation}}
The starting point is the Fierz transformation, which  brings $b \bar b$ together~\cite{Ali:2019roi}: 
\bea
&& {\cal T}(J=0)= \left| \left( b^{}_{}  b^{}_{} \right)_{\bar{3}}^{\;1}  \left( \bar{b}^{}_{} \bar{b}^{}_{} \right)_{3}^{\;1}    \right \rangle_{1}^{\;0}=-\frac{1}{2}\left(    \sqrt{\frac{1}{3}} \left| \left( b^{} \bar{b}^{} \right)^{\;1}_{1}   \left( b^{}  \bar{b}^{} \right)^{\;1}_{1} \right\rangle^{\;0}_{1}   -\sqrt{\frac{2}{3}} \left| \left( b^{} \bar{b}^{} \right)^{\;1}_{8}  \left( b^{}  \bar{b}^{} \right)^{\;1}_{8}   \right\rangle^{\;0}_{1}  \right) + \nonumber \\ 
&&+\frac{\sqrt{3}}{2} \left(    \sqrt{\frac{1}{3}}  \left| \left( b^{} \bar{b}^{} \right)^{\;0}_{1}   \left( b^{}  \bar{b}^{} \right)^{\;0}_{1} \right\rangle^{\;0}_{1}  -\sqrt{\frac{2}{3}} \left| \left( b^{}  \bar{b}^{} \right)^{\;0}_{8}  \left( b^{}  \bar{b}^{} \right)^{\;0}_{8}   \right\rangle^{\;0}_{1}   \right).
 \label{fierz1} 
 \eea
quark bilinears are normalised to unity, subscripts denote the dimension of colour representations, and superscripts the total spin.%  the configuration $\left( b^{} \bar{b}^{} \right)^{\;1}_{1}  $  and $\left( b^{} \bar{b}^{} \right)^{\;0}_{1} $ are colour singlets with the quantum number of the $\Upsilon$ and $\eta_b$ mesons, respectively. 
~Similarly, for the $J=2$ tetraquark, one finds:
\bea
&& {\cal T}(J=2)= \left| \left( b^{}_{}  b^{}_{} \right)_{\bar{3}}^{\;1}  \left( \bar{b}^{}_{} \bar{b}^{}_{} \right)_{3}^{\;1}    \right \rangle_{1}^{\;2} =\left(    \sqrt{\frac{1}{3}}  \left| \left( b^{} \bar{b}^{} \right)^{\;1}_{1}   \left( b^{}  \bar{b}^{} \right)^{\;1}_{1} \right\rangle^{\;2}_{1}  -\sqrt{\frac{2}{3}} \left| \left( b^{}  \bar{b}^{} \right)^{\;1}_{8}  \left( b^{}  \bar{b}^{} \right)^{\;1}_{8}   \right\rangle^{\;2}_{1}   \right).
 \label{fierz2} 
 \eea

We describe the ${\cal T}$ decay as due to individual decays into lower mass states of one of the $b \bar b$ pairs in \eqref{fierz1}, described as follows. 
\begin{enumerate}
\item The colour singlet, spin $0$ pair decays into  $2$  gluons,  which are converted  into confined, light hadrons (i.e. not containing $b$ flavour) with a rate  of the  order of $\alpha_S^2$; taking the spectator $b\bar b$ pair into account,  this decay leads to: ${\cal T} \to \eta_b+{\rm light~ hadrons}$.
\item The colour singlet, spin $1$ pair decays into $3$  gluons,  which are converted  into confined light hadrons  with a rate  of the  order of  $\alpha_S^3$, leading to:  ${\cal T} \to \Upsilon+{\rm light ~hadrons}$; final states $\Upsilon +\mu^+ \mu^-$ and $4\mu$ are also produced.
\item  The colour octet, spin $1$ pairs annihilate into one gluon, which materializes into a pair of light quark flavours, $q=u,d,s,c$, the latter recombine with the spectator pair to produce a pair of  lower-lying, open-beauty mesons $B_q\bar B_q$ and $B^*_q\bar B^*_q$, with a rate  of the  order of $\alpha_S^2$. 
\item  The colour octet, spin $0$  pairs annihilate into a pair of lighter quarks (necessary to neutralize the colour of the spectator $b\bar b$ pair) with amplitude of the  order of  $\alpha_S^2$ and  with a  rate  of the order of $\alpha_S^4$, which we neglect. % {\colour{red} {\bf CHECK CHECK}}
\end{enumerate}

The total $\cal T$ decay is the sum of these individual decay rates, which are obtained from the simple formula~\cite{landlif}
\be
\Gamma((b\bar b)_c^s)=|\Psi(0)_{\cal T}|^2 v\sigma ((b\bar b)_c^s \to f)
\label{master1}
\ee
$|\Psi(0)_{\cal T}|^2$ is the overlap probability of the annihilating pair, $v$ the relative velocity and $\sigma$ the spin-averaged annihilation cross section in the final state $f$ \footnote{Our method of calculation is borrowed from the theory of $K$ electron capture, where an atomic electron reacts with a proton in the nucleus to give a final nucleus and a neutrino. The rate is computed from formula \eqref{master1}, in which $|\psi(0)|^2$  is  the overlap probability of the electron to the proton in the nucleus and $\sigma$ the electron-proton weak cross section. The other electrons in the atom act as spectators, and rearrange later into a stable atom, after emission of radiation, with unit probability.}.
The spectator $b \bar b$ pair, given the lack of extra energy, appears as $\eta_b$ or $\Upsilon$ on the mass shell, or  combines with the outgoing $q\bar q$ pair into an open-beauty meson pair.  Our results are valid in the situation where the tetraquak mass is just below the $2\Upsilon$ threshold and each non-relativistic pair has mass very  close to $2m_b$.%, the sum of $b$ and $\bar b$ quark masses.

%\emph{{\bf Overlap probabilities.}}
We normalise the overlap probabilities to that of $|\Psi_\Upsilon(0)|^2$, which can be derived from the $\Upsilon$ decay rate into lepton pairs.
Eq.~\eqref{master1} applied to this case gives:
\be
\Gamma(\Upsilon \to \mu^+ \mu^-)=Q_b^2 ~\frac{4\pi \alpha^2}{3}  \frac{4}{m_\Upsilon^2}~ |\Psi_\Upsilon(0)|^2, ~(Q_b=-1/3). \label{QMpict}
\ee
% It is convenient to connect with the  Vector Meson Dominance parameter, defined by~\cite{Schildknecht:2005xr}\beJ^\mu(x) =\bar b(x) \gamma^\mu b(x)=\frac {m_\Upsilon^2}{f}~ \Upsilon^\mu (x)\ee$f$ being a phenomenological parameter, a pure number. With this relation, one obtains again the $\mu^+\mu^-$ rate as\bea&&\Gamma(\Upsilon \to \mu^+ \mu^-)=\frac{4\pi\alpha^2}{27}\frac{m_\Upsilon}{f^2}.\label{VMD}\eeaand:\be |\Psi_\Upsilon(0)|^2=\frac{m_\Upsilon^{3}}{4f^2} \label{fandpsi}\ee
 It is useful to connect with the  Vector Meson Dominance parameter~\cite{Schildknecht:2005xr} defined by
\be
J^\mu(x) =\bar b(x) \gamma^\mu b(x)=\frac{m_\Upsilon^2}{f}~\Upsilon^\mu (x)
\ee
$f$ being a pure number. %With this relation, 
One obtains~\cite{pdg}
\bea
%&&\Gamma(\Upsilon \to \mu^+ \mu^-)=\frac{4\pi\alpha^2}{27}\frac{m_\Upsilon}{f^2}\label{VMD}\\
&&|\Psi_\Upsilon(0)|^2=\frac{m_\Upsilon^{3}}{4f^2}; \label{fandpsi} \;\; f=13.2;~|\Psi_\Upsilon(0)|^2\sim 1.2~{\rm GeV}^{3}.\label{fandpsi2}
\eea

\emph{{\bf Numerical results.}} From Eq.~\eqref{master1},  the contribution to the ${\cal T}$ decay rate of the colour singlet, spin $0$ decay  is
\bea
&&\Gamma_0=\Gamma( {\cal T}\to \eta_b + {\rm light ~hadrons})= 2\cdot \frac{1}{4} \cdot  |\Psi(0)_{\cal T}|^2 v\sigma ((b\bar b)_1^0\to 2 ~{\rm gluons})\notag \\
&&= \frac{1}{2}~\Gamma(\eta_b)\cdot \xi= 5~{\rm MeV}\cdot \xi
\eea
the factor $2$ arises because of the two  $(b \bar b)_1^0$ pairs, $1/4$ is the spectroscopic coefficient %of the colour singlet spin $0$ term 
in \eqref{fierz1} and we have approximated
\be
|\Psi(0)_\Upsilon|^2 v\sigma ((b\bar b)_1^0\to 2 ~gluons)\sim \Gamma(\eta_b)=10~{\rm MeV}.
\ee

In a similar way, we obtain
\bea
&&\Gamma_1=\Gamma( {\cal T}\to \Upsilon + {\rm light ~hadrons})= 2\cdot \frac{1}{12} \cdot  |\Psi(0)_{\cal T}|^2 v\sigma ((b\bar b)_1^1\to 3 ~{\rm gluons})=\notag \\
&&= \frac{1}{6}~\Gamma(\Upsilon)\cdot \xi= 9~{\rm keV}\cdot \xi \notag \\
&&\Gamma_2=\Gamma( {\cal T}\to \Upsilon + \mu^+\mu^-)= B_{\mu\mu} \Gamma_1= 0.22~{\rm keV}\cdot \xi\notag \\
&& \Gamma_4=\Gamma( {\cal T}\to 4\mu)= B_{\mu\mu} ^2\Gamma_1= 5.5~10^{-3}~{\rm keV}\cdot \xi \label{gam4mu}
\eea

%@@@@@@@@@@@@@@@@@@@@@@@
 \begin{figure}[htbp]
   \centering
   \includegraphics[width=0.6 \linewidth]{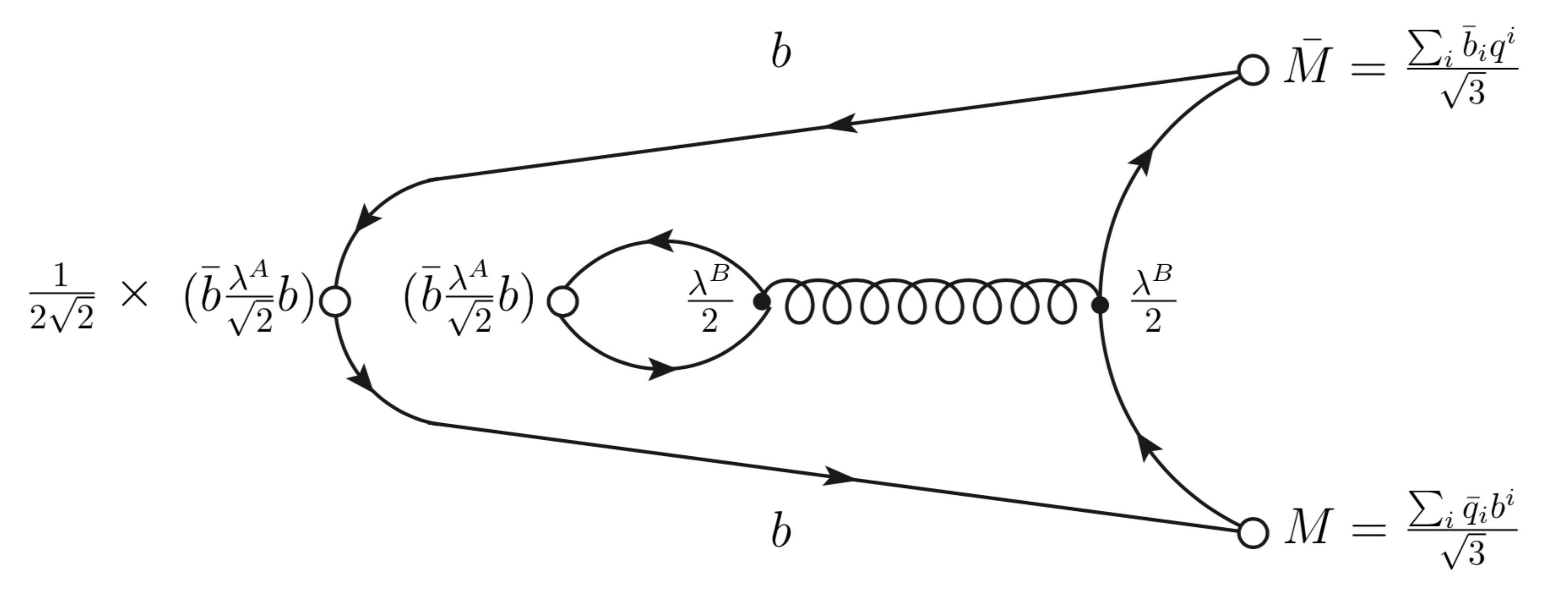}
   \caption{Colour flow in $b\bar b$ annihilation.}
         \label{chanpaton}
\end{figure}
%@@@@@@@@@@@@@@@@@@@@@
Finally, we consider the annihilation of the $(b \bar b)^1_8$ in light quark pairs. This is illustrated in Fig.~\ref{chanpaton}. Open circles represent the insertion of quark bilinears, and black dots the QCD vertices. Colour matrices and normalizations are explicitely indicated. The numerical factor associated to the traces of the colour matrices along fermion closed paths, C (the Chan-Paton factor \cite{Paton:1969je}) gives the effective coupling constant of the process,
$
\alpha_{eff}=   C \alpha_S, 
$
which is what replaces $Q_b \alpha$ in Eq.~\eqref{QMpict}. From Fig.~\ref{chanpaton} we read: $C=\sqrt{2}/3$ and we find:
\bea
&& \Gamma_5=\Gamma( {\cal T}\to M(b\bar q)+M(q\bar b))=
 2\cdot \frac{1}{6}\cdot \frac{2}{9}\cdot \left(\frac{4\pi\alpha_S^2}{3}  \frac{4}{m_\Upsilon^2}\right)~|\Psi(0)_\Upsilon|^2\cdot \xi \label{gammames}
\eea
The factor $2$ arises from the two choices of the annihilating bilinear\footnote{given the symmetry of the tetraquark, we may call $b_1$ the annihilating $b$ quark and pair it, in Eq.~\eqref{fierz1}, to either $\bar b_1$ or $\bar b_2$.}. We have inserted the spectroscopic factor of the spin $1$ colour octet from~\eqref{fierz1} and the Chan Paton factor. In parenthesis $v\sigma(b\bar b\to q\bar q)$. Using Eq.~\eqref{fandpsi}, $\alpha_S=0.2$ and massless $q$,  we obtain
\be
 \Gamma_5= \frac{8\pi}{81}\left(\frac{\alpha_S}{f}\right)^2m_\Upsilon\cdot\xi= 0.67~{\rm MeV} \cdot \xi
 \label{gammames2}
\ee
and 
\be
\Gamma({\cal T})= \Gamma_0+\Gamma_1 + 4\Gamma_5 =7.7~{\rm MeV} \cdot \xi
\label{gamtot}
\ee
A non-vanishing mass of the final quark brings a negligible correction even for  the charm.
Eq.~\eqref{gammames2} refers to the total decay rate into pseudoscalar and vector meson pairs. 
We can separate the two rates according to the following argument.
In non-relativistic notation, the spin-colour structure of the final state after annihilation corresponds  to the operator ($\cal O$ is  normalized to unit norm, see~\cite{Ali:2019roi}):
 \be
 {\cal O}_{fin}=\frac{1}{4\sqrt{2}} \sum_A~\left(b_C \lambda^A\sigma_2{\pmb \sigma}b\right)\cdot \left(q_C\lambda^A \sigma_2{\pmb \sigma}q \right)
 \ee
 Using the appropriate Fierz-rearranging relations, one sees that\footnote{Fierzing colours produces $b\bar q$ and $q \bar b$ bilinears in colour singlets and colour octet; one may argue that gluons from the vacuum will screen colour octet charges~\cite{Bali:2000gf,Maiani:2019lpu}.} $B_q B^*_q$ and ${\pmb B}_q {{\pmb B^*}}_q$ pairs are produced in the spin combination,
 $
 \frac{1}{2}\left[ (B_q B^*_q) +({\pmb B}_q\cdot  {{\pmb B^*}}_q)\right]
 $
 and the rate in $\bar b q+\bar q b$ is shared between pseudoscalar and vector mesons in the ratio $1:3$.

\section{{\bf The value of $|\Psi_{\cal T}(0)|^2$}}
 A value for $|\Psi_{\cal T}(0)|^2$ can be obtained from the calculation in \cite{Anwar:2017toa,Liu}. Constituent coordinates are defined as
  \bea
 && {\bf x},~{\bf y}:~{\rm antiquarks};  {\bf z},~{\bf 0}:~{\rm quarks}\notag
 \eea
 One defines the Jacobi coordinates
 \be
 {\pmb \xi}_1= {\pmb x}-{\pmb y};~{\pmb \xi}_2= {\pmb z};~{\pmb \xi}_3= {\pmb x}+{\pmb y}-({\pmb z}+{\pmb 0})  \label{tetracoord}
 \ee 
 The ${\cal T}$ wave function is a product of normalized gaussians with parameters  $\beta_1=\beta_2=0.77~{\rm GeV}~{\rm },~\beta_3=0.60~~{\rm GeV}$, obtained by minimising the expectation of the Hamiltonian 
 of Ref. \cite{Anwar:2017toa,Liu}. The equality $\beta_1=\beta_2$ is due to Charge conjugation invariance.
 By elementary integrations, one can obtain the wave function squared in the variable $\bf x$, i. e. the separation of an antiquark from the quark in the origin, or in the variable 
$ 
 {\pmb \ell}= \frac{1}{2} {\pmb \xi}_3, \label{cograv}
$ 
 i.e. the separation of the centers of gravity of quarks and antiquarks. 
 
  %@@@@@@@@@@@@@@@@@@@@@@
  \begin{table}[htb!]
\centering
    \begin{tabular}{|c|c|c|c|}
     \hline
 {\footnotesize Particle}&{\footnotesize Method} & {\footnotesize $|\Psi(0)|^2~({\rm GeV}^3)$} & {\footnotesize $ \sqrt{<R^2>}$~(fm)}\\ \hline 
 {\footnotesize ${\Upsilon}$}&{\footnotesize Eq.~\eqref{fandpsi2}}&{\footnotesize $1.2$} &{\footnotesize $0.13$, using Eq.~\eqref{gaussrel}}\\ \hline
{\footnotesize ${\Upsilon}$}&{\footnotesize Eq.~\eqref{ourgaussian} }&{\footnotesize $0.16$} &{\footnotesize $0.25$}\\ \hline
{\footnotesize ${\Upsilon}$}&{\footnotesize Ref.~\cite{Liu}}&{\footnotesize $0.24$ (gaussian w.f.)} &{\footnotesize $0.22$}\\ \hline 
{\footnotesize ${\cal T}$}&{\footnotesize ${\pmb x}$,~Eq.~\eqref{tetracoord}}&{\footnotesize $0.094$}  &{\footnotesize $0.30$} \\ \hline
{\footnotesize ${\cal T}$}&{\footnotesize ${\pmb \ell}$,~Eq.~\eqref{cograv}}&{\footnotesize $0.31$}  &{\footnotesize $0.20$}\\ \hline 
\end{tabular}
 \caption{\footnotesize{ Estimates of overlap probability and radius, for $\Upsilon$ and $\cal T$.}}
\label{due}
\end{table}  
  %@@@@@@@@@@@@@@@@@@@@
One finds
\bea
&&|\Psi_{\cal T}({\pmb x})|^2 =(\frac{\gamma}{ \pi})^{3/2}  \cdot e^{-\gamma{\pmb x}^2},~\sqrt{\gamma}=\sqrt{\left(\frac{4\beta_1^2 \beta_3^2}{\beta_1^2+2\beta_3^2}\right)}=0.81~{\rm GeV} ;\label{ics} \\
&&|\Psi_{\cal T}({\pmb \ell})|^2 =(\frac{(2\beta_3)^2}{ \pi})^{3/2}  \cdot e^{-4\beta_3^2{\pmb \ell}^2} \label{ell}
\eea
For gaussian wave function there is a fixed relation 
$ 
|\Psi(0)|^2=(\frac{3}{2\pi R^2})^{3/2}. \label{gaussrel}
$
%and  overlap probabilities scale with the inverse cube of the radius. 

To compute $|\Psi_{\Upsilon}(0)|^2$, we  follow Ref.~\cite{Anwar:2017toa}.  We obtain the wave function
\bea 
&&|\Psi_\Upsilon({\pmb x})|^2 =(\frac{\beta_\Upsilon^2}{ \pi})^{3/2}  \cdot e^{-\beta_\Upsilon^2{\pmb x}^2},~\beta_\Upsilon=0.96~{\rm GeV}  \\
&&|\Psi_\Upsilon(0)|^2 =0.159~{\rm GeV}^3;~R_\Upsilon=\sqrt{\frac{3}{2\beta_\Upsilon}}= 0.252~{\rm fm}. \label{ourgaussian}
\eea
Tab.~\ref{due}  reports the results obtained from different methods. 
As the table shows, the overlap probability from $\Upsilon$ leptonic decay is substantially larger than the one obtained in the gaussian model, which, for the radius, agrees reasonably with the independent evaluation of Ref.~\cite{Liu}. The discrepancy underlines the need to estimate $\xi= |\Psi_{\cal T}(0)|^2/|\Psi_{\Upsilon}(0)|^2$  by means of the same method for the numerator and denominator. 

With the two definitions of the radius in Eqs.~\eqref{ics} and \eqref{ell}, for $\cal T$, and with Eq.~\eqref{ourgaussian} for $\Upsilon$, we find
$
\xi(\pmb x)=0.58< \xi <1.95=\xi({\pmb \ell}).
$
A good compromise is the geometrical mean, with the previous result used as an error estimate:
\be
\xi_{\rm th}=\sqrt{\xi(\pmb x)\xi({\pmb \ell})}= 1.1^{+0.9}_{-0.5} \label{bestguess}
\ee

Branching ratios do not depend on $\xi$ and are not affected by this error.
%@@@@@@@@@@@@@@@@@@@@@@@@@@@@@@@@@@@@@@@@@@@@
%@@@@@@@@@@@@@@@@@@@@@@@@@@@@@@@@@@@@@@
\section{{\bf Tetraquark cross section in the 4 $\mu$ and $B^{(*)}_f B^{(*)}_f$  channels}}

By  combining Eqs.~\eqref{gam4mu} and \eqref{gamtot} we obtain a very low branching fraction for ${\cal T}\to 4 \mu$:
\be
B_{4\mu}=B({\cal T}\to 4\mu)= 7.2~10^{-7}\label{b4mu}
\ee

The cross section  upper-limit obtained from \eqref{b4mu} is
 \bea
 &&\sigma_{theo.}({\cal T}\to 4 \mu) %= \sigma(pp \to {\cal T})  B_{4\mu}
  \leq \sigma(pp \to 2 \Upsilon) B_{4\mu}= 
\left\{ \begin{array}{c}
 0.049  ~\text{fb, with $\sigma(pp \to 2 \Upsilon)\simeq 69$ pb  \cite{CMS2Upsilon}}\\
 0.056 ~\text{fb, with $\sigma(pp \to 2 \Upsilon)\simeq 79$ pb  \cite{Sirunyan:2020txn}}
 \end{array}
 \right.
\label{uplim}
 \eea
We observe that $\sigma(pp \to 2 \Upsilon)\simeq 69(79)$ pb is the  two-$\Upsilon$ production cross section
measured by CMS at 8 TeV \cite{CMS2Upsilon} (13 TeV \cite{Sirunyan:2020txn}). 

 According to \eqref{uplim}, the  four-muon tetraquark cross section  is far below the  current LHCb capabilities, the upper limit  in Eq.\eqref{uplim} being more than
 350 times lower than  the CL of the $95 \%$ LHCb  upper limit of  $\simeq 20$ fb quoted in Eq.~\eqref{uplhcb}.

Summing over the four light flavours, we see from Tab.~\ref{uno} that  decays into meson pairs account for about $35\%$ of $\cal T$ decays.
Decay into $B^+ B^-$ mesons may provide a promising channel to discover the $4b$ tetraquark. 
In Tab.  \ref{uno}  we report the tetraquark open-bottom  branching fractions and the upper limits to the  tetraquark  two-lepton cross section, 
  $\sigma_{theo.}({\cal T} \to 2 B_q^{}\to  2 l)$, calculated as
   \bea
&& \sigma_{theo.}({\cal T} \to 2 B_f^{}\to  2 l) = \sigma(pp \to {\cal T})  Br({\cal T} \to 2 B_q)~[Br(B_{q}\to l+\bar{\nu}+X)]^2
 \nonumber\\
&& \leq \sigma(pp \to 2 \Upsilon) Br({\cal T} \to 2 B_q)~[Br(B_{q}\to l+\bar{\nu}+X)]^2
\label{Spred}
 \eea
Decays such as $B_f^{*}\to  l+ \bar{\nu}+X$ occur by means of intermediate electromagnetic  decays, for example $B_f^{*}\to B_f^{}+\gamma \to  l+ \bar{\nu}+X +\gamma$.%, and so their cross section is much less than the  $B_f^{}\to  2 l$ cross section. 
~For this reason, the excited open-bottom channels are not  suited to  the search for tetraquark states.
%With $\sigma(pp \to 2 \Upsilon)=68.8$ pb~\cite{CMS2Upsilon}, upper limits to the cross sections 
Upper limits for ${\cal T}$ production and decay into into  $B^{+} B^{-}, B^0 \bar{B}^0$ and $B_s^0 \bar{B}_s^0$ are reported in Tab.~\ref{tab:width}.

 \renewcommand{\arraystretch}{1.2}
 \begin{table}[htbp]
  \centering
\caption{Upper limits  of two- and four-lepton cross sections via ${\cal T}$ production calculated using as inputs 
the  two-$\Upsilon$ production cross sections measured by CMS at 8 TeV \cite{CMS2Upsilon} and 13 TeV  \cite{Sirunyan:2020txn}.
% {\color{red}  
%In the last column our … using as input the sigma from 27 in the last
%Column the first number is the predicate 2 lepton vc supper limit calculated using as input sigma from ref. 27 
%$\sigma(pp \to 2 \Upsilon)\simeq 69(79)$ pb is the two-$\Upsilon$ production cross section
%measured by CMS at 8 TeV \cite{CMS2Upsilon} (13 TeV \cite{Sirunyan:2020txn}). } 
}
% \begin{tabular}{ccccc}
  \begin{tabular}{*6c}
\hline
\hline
Decay Channel & Predicted &&& \multicolumn{2}{c}{Predicted  two-lepton}   \\
&BF&&& \multicolumn{2}{c}{cross section  upper limit (fb)}     \\
&&&&8 TeV& 13 TeV\\
\hline
${\cal T}(J=0) \to B^{+} B^{-} (B^0 \bar{B}^0, B_s^0 \bar{B}_s^0) \to 2\ell+\dots$ & 0.022  &&& 34 & 39\\
%${\cal T} \to  B^{0}\bar{B}^{0}$ &  0.05  & 34 \\
%${\cal T} \to  B^{*+}B^{*-}$ & 7.84 &  - \\
%${\cal T} \to  B^{*0}\bar{B}^{*0}$ & 7.84 & -  \\
%${\cal T} \to  B^{0}_c \bar{B}^{0}_c \to 2\ell+\dots$ & 0.045 & 38   \\
${\cal T}(J=0) \to  4\mu $ & $7.2\cdot 10^{-7}$ &&& $0.049$ &  0.056  \\
${\cal T}(J=2) \to  4\mu $ & $20\cdot 10^{-7}$ &&&  $0.14$ & 0.16  \\
%${\cal T} \to  B^{+}_c \bar{B}^{-}_c $ & 7.17&  $1.23\; 10^{-6}$\\
%${\cal T} \to  B^{*+}_c \bar{B}^{*-}_c $ &  7.47& - \\
\hline
\hline
\end{tabular}
\label{tab:width}
\end{table}
 In conclusion, we propose the $B^{+}B^{-}$, $B^{0}\bar{B}^{0}$ and $B^{0}_s\bar{B}^{0}_s$  channels   in the search for full-bottom tetraquarks in future LHCb upgrades. The $4\mu$ signal produced by the $J^{PC}=2^{++}$ tetraquark may not be hopeless.

%  \tableofcontents

\end{document}